\documentclass{aa}
\usepackage{graphicx}
\usepackage{hyperref}
\usepackage{natbib}
\usepackage{float}
\usepackage{longtable}
\usepackage{csvsimple}
\usepackage{color}
\usepackage{textcomp}
\usepackage{caption}
\usepackage{subfigure}
\usepackage[retainorgcmds]{IEEEtrantools}
\usepackage[utf8]{inputenc}
\usepackage[varg]{txfonts}
\usepackage{mathabx}

\usepackage{hyperref}
\hypersetup{
    colorlinks = true,
    linkcolor = {blue}, 
    citecolor = {blue},
    urlcolor = {blue}
}

\begin{document}
    \title{Weak evidence for variable occultation depth of 55 Cnc e with TESS}
    
    \titlerunning{Variable occultation depth of 55 Cnc e with TESS}
    \author{E.A.~Meier Vald\'es
        \inst{1}
        \and 
        B.M.~Morris
        \inst{1}
        \and 
        R.D.~Wells
        \inst{1}
        \and 
        N.~Schanche
        \inst{1}
        \and
        B.-O.~Demory
        \inst{1}}
    \institute{Center for Space and Habitability (CSH), University of Bern,
              Gesellschaftsstrasse 6, 3012 Bern, Switzerland\\
              \email{erik.meiervaldes@unibe.ch}
              }
\date{Received 12 April 2022 / Accepted 14 May 2022}

\abstract
{55 Cnc e is in a 0.73 day orbit transiting a Sun-like star. It has been observed that the occultation depth of this Super-Earth, with a mass of 8$M_{\bigoplus}$ and radius of 2$R_{\bigoplus}$, changes significantly over time at mid-infrared wavelengths. Observations with Spitzer measured a change in its day-side brightness temperature of 1200 K, possibly driven by volcanic activity, magnetic star-planet interaction, or the presence of a circumstellar torus of dust.}
{Previous evidence for the variability in occultation was in the infrared range. Here we aim to explore if the variability exists also in the optical.}
{TESS observed 55 Cnc during sectors 21, 44 and 46. We carefully detrend the data and fit a transit and occultation model for each sector in a Markov Chain Monte Carlo routine. In a later stage we use the Leave-One-Out Cross-Validation statistic to compare with a model of constant occultation for the complete set and a model with no occultation.}
{We report an occultation depth of 8$\pm$2.5 ppm for the complete set of TESS observations. In particular, we measured a depth of 15$\pm$4 ppm for sector 21, while for sector 44 we detect no occultation. In sector 46 we measure a weak occultation of 8$\pm$5 ppm. The occultation depth varies from one sector to the next between 1.6 and 3.4 $\sigma$ significance. We derive the possible contribution on reflected light and thermal emission, setting an upper limit on the geometric albedo. Based on our model comparison the presence of an occultation is favoured considerably over no occultation, where the model with varying occultation across sectors takes most of the statistical weight.}
{Our analysis confirms a detection of the occultation in TESS. Moreover, our results weakly lean towards a varying occultation depth between each sector, while the transit depth is constant across visits.}


\maketitle

\section{Introduction}
\label{section:introduction}

55 Cnc e was first discovered by \citet{McArthur_2004} via radial velocity (RV) observations with the Hobby-Eberly Telescope (HET) in a 2.808 day orbit and later found to be an alias of its true period of 0.7365 days \citep{Dawson_2010}. \citet{Winn_2011} and \citet{Demory_2011} confirmed the period and detected the planet to be transiting its host star, one of the brightest stars (V=6.0) known to host planets. 

The conundrum of 55 Cnc e's nature began with the detection of a phase modulation that was too large to be caused by reflected starlight and thermal emission of the planet \citep{Winn_2011}, later to be found varying over time \citep{Dragomir_2012, Sulis_2019}. Given the short separation to the star, a possible explanation is star-planet interaction. \citet{Folsom_2020} derived a map of the large-scale stellar magnetic field of 55 Cnc, concluding that planet e orbits within the Alfvén surface of the stellar wind, allowing for magnetic star-planet interactions. 

\citet{Demory_2015} found a 300\% difference in occultation depth between 2012 and 2013 in the Spitzer/IRAC \citep{Werner_2004, Fazio_2004} 4.5 $\mu$m channel, which translates into a change in day-side brightness temperature of approximately 1200 K, later confirmed independently by \citet{Tamburo_2018}. This could be caused by volatile loss through surface evaporation, volcanic activity on the surface of the planet or the presence of an inhomogeneous circumstellar torus of dust \citep{Demory_2015, Tamburo_2018, Sulis_2019}.

Observations over different wavelengths can shed light on the nature of a planet, providing complementary information about the planetary atmosphere. This system has already benefited from observations in the IR \citep{Demory_2011, Demory_2015, Demory_2016b, Tamburo_2018}, Optical \citep{Winn_2011, Dragomir_2012, Sulis_2019,Kipping_2020,Morris_2021}, Far-UV \citep{Bourrier_2018b} and X-ray \citep{Ehrenreich_2012}. This list is not exhaustive. Here we present the analysis of the transit and occultation for all observations made by TESS \citep{Ricker_2015} so far.    

\section{Methods}
\label{section:method}

\subsection{TESS observations}
\label{section:TESS}

55 Cnc e (TIC~332064670) was observed by TESS during sector 21, 44 and 46. Each sector consists of two TESS orbits. The time interval between the first set of observations and the second is approximately 600 days. The gap between the observations of sector 44 and 46 consists of 29 days. The target was not observed during sector 45. The observations include a total of 93 transits and occultations each. We use the 120-second cadence Pre-search Data Conditioning (PDC) lightcurve data from the Science Processing Operations Center (SPOC) pipeline \citep{Jenkins_2016}.

To prepare our data, first we remove all points above 4$\sigma$ from the median of the absolute deviations. We compute a Lomb-Scargle periodogram \citep[][and references therein]{VanderPlas_2018} to check for significant periodicities. Besides planet e's period and aliases, there is a strong signal between 6 and 6.5 days, corresponding to momentum dumps. To remove trends in the data we mask all transits and occultations, then fit a robust M-estimator using Tuckey's biweight function implemented in \texttt{wotan} \citep{Hippke_2019}, setting the length of the filter window matching planet e's orbital period. After detrending we do a second clipping to remove outliers above 4$\sigma$.

Although the PDC lightcurves were already corrected for background noise, stray light and several other quality flags, we notice flux ramps before or after momentum dumps, which often coincide with stray light reflected from Earth. Since these short-timescale events are difficult to correct without affecting astrophysical signals, we follow a similar procedure as \citet{Beatty_2020} and trim a portion of the data preceding and following these events. In particular, we remove 1.42 days at the beginning, 0.06 days at the end of the first orbit and 1.09 days at the beginning of the second orbit in sector 21; 0.8 days at the beginning, 0.02 days at the end of the first orbit and 2.43 days at the beginning of the second orbit in sector 44; 2.261 days and 2.43 days at beginnings of both orbits in sector 46. 
The information regarding momentum dumps, quality flags and summary of each sector is obtained from the TESS Data Release notes\footnote{\url{https://archive.stsci.edu/tess/tess_drn.html}} for each sector and the corresponding Data Validation (DV) files. The photon noise contribution for 55 Cnc (TESS mag = 5.2058) over a 2-hour timescale is 63.26 ppm, 57.94 ppm and 60.69 ppm for sector 21, 44 and 46, respectively.

We also remove all flagged data points from the time series. In total, we remove 1678 of 17319 data points for sector 21, 647 of 15247 data points for sector 44 and 4097 of 16714 data points for sector 46. After this process, our photometric data contains 84 transit and 84 occultations. 

\subsection{Lightcurve analysis}
\label{section:lightcurve}

First, we restrict the dataset to 0.25 in phase before and after mid-transit to ensure we cover more than twice the transit duration (0.0648 days, \citealt{Sulis_2019}) preceding and following epoch of mid-transit. Keeping the transits and occultations masked, we compute the out-of-transit mean flux for each segment containing one of the 84 transits and then detrend the observations. The reasons for doing this step are twofold: first, to ensure a normalised out-of-transit mean flux of unity and to keep our light curve model MCMC with as few free parameters as possible.

The lightcurve model is based on those of \citet{Mandel_2002}, implemented in the \texttt{exoplanet} Python package \citep{Foreman-Mackey_2021}. All three sectors are analysed together. We assume a circular orbit \citep{Bourrier_2018} and a quadratic stellar limb-darkening law. The priors on the limb darkening coefficients are obtained from a list \citep{Claret_2017} of coefficients for TESS based on a 1-D Kurucz ATLAS stellar atmosphere model \citep{Castelli_2004}. In our transit model we fit for the time of mid-transit, orbital period, quadratic limb darkening coefficients, planet-to-star radius ratio and impact parameter. Our model is implemented in a Markov Chain Monte Carlo (MCMC) with the PyMC3 probabilistic programming package \citep{Salvatier_2016}. We fit for a planet-to-star radius ratio for each sector, while the rest of parameters represent a single value for all sectors. In this manner, we can compare the transit and occultation depth between sectors instead of obtaining a composite fit. To compute the transit depth for a given stellar limb darkening law and impact parameter, we implement the analytic solutions from \citet{Heller_2019} in our MCMC algorithm.

The second step consists of freezing the best-fit parameters from the transit model \citep{Garhart_2020} for another MCMC run, fitting for the occultation depth. In this case, the limb-darkening coefficients are fixed to zero. To fit for the occultation, we consider data points before and after 0.25 in phase from the occultation centre. Here, we allow the occultation depth parameter to explore negative values. For both the transit and occultation model MCMC, we check that the chains are well mixed and that the Gelman-Rubin statistic is below 1.01 for all parameters \citep{Gelman_1992}. Finally, we compute a power spectrum of the residuals to make sure that after removing the signal of the planet, there are no significant signals remaining.

We also construct an MCMC algorithm to fit a single occultation depth parameter on the complete observations and a model with the occultation depth fixed to zero. 

To ensure our models are robust, after running each model, we estimate the out-of-sample predictive accuracy with Leave-One-Out Cross-validation \citep{Vehtari_2016} to detect any data point with a shape parameter $\hat{k}$ of the Pareto distribution greater than 0.7. Essentially, if a single point has a shape parameter greater than 0.7, the model is considered unreliable \citep{Vehtari_2015}. In total, we reject 22 points after 5 iterations. 

\section{Results}
\label{section:results}

\subsection{Transit depth}
\label{subsection:transit}

\begin{figure}
    \centering
    \resizebox{\hsize}{!}{\includegraphics{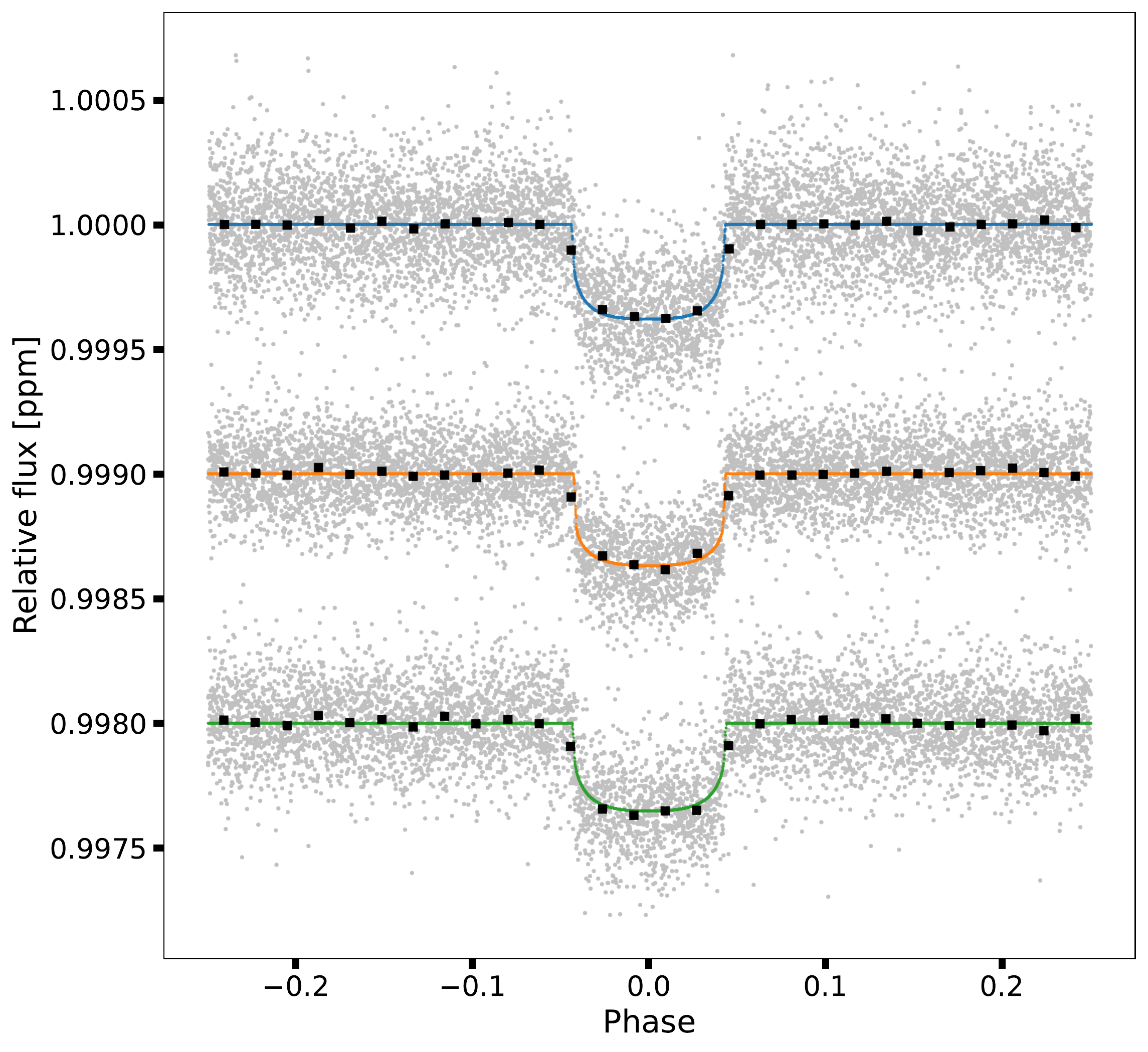}}
    \caption{Phase folded relative flux displaying the transit. The lightcurves are shifted down for clarity. The top lightcurve corresponds to sector 21, middle lightcurve to sector 44 and at the bottom to sector 46. Silver points are detrended flux measurements. The continuous lines show the transit model, where sector 21 is depicted in blue, sector 44 in orange and sector 46 in green. We will keep this convention for the rest of this work. Black squares represent binned data. Note that bins are for visualization and were not used in our analysis.}
    \label{fig:transit}
\end{figure}

In Table \ref{tab:bestfit}, we present best-fit values. We find a transit depth consistent for all sectors within the uncertainties. Figure \ref{fig:transit} shows a portion of the phase including the transit for each sector and the corresponding transit model overlapped. Figure \ref{fig:transit corner} in Appendix \ref{appendix:corner} shows the posterior distribution and correlations between all parameters sampled with our MCMC. The unbinned residual Root Mean Square (RMS) is 166.2 ppm, 134.02 ppm and 146.48 ppm for sector 21, 44 and 46, respectively (see Appendix \ref{appendix:rms}).

Based on the marginal 1.6$\sigma$ difference, we conclude that there is no variability in the transit depth during the time of observation in the TESS bandpass. Compared to the observations done by CHEOPS and analysed by \citet{Morris_2021}, their best-fit values imply a similar transit depth of $339.72_{-23.54}^{25.33}$. \citet{Winn_2011} reported a transit depth of $380 \pm 52$ ppm for MOST, while \citet{Tamburo_2018} and \citet{Demory_2015} obtained $336 \pm 18$ and $360 \pm 26$ for Spitzer, respectively.

\begin{table}[h]
        \centering\setstretch{1.5}
        \caption{Best-fit parameters for all sectors. The priors on stellar radius $r_{star}$ and stellar mass $m_{star}$ are based on \citet{VonBraun_2011}. The value presented is the median and 1-$\sigma$ confidence interval. The symbols stand for: $P$, period; $T_{0}$, BJD mid-transit time; $u_{0}$ and $u_{1}$, quadratic limb-darkening coefficients; $R_{p}/R_{s}$, planet-to-star radius ratio; $\delta_{t}$, transit depth and $\delta_{e}$, occultation depth. The number in the subscript refers to the corresponding sector.}
        \begin{tabular}{c c}
            \hline
            \hline
            Priors &  \\
            \hline 
             $r_{star}$ [$R_{\odot}$] & $0.943_{-0.01}^{+0.01}$ \\
             
             $m_{star}$ [$M_{\odot}$] & $0.905_{-0.015}^{+0.015}$ \\
             
            \hline
             Parameter & Value \\
            \hline
            
             $P$ [days] & $0.73654627_{-0.00000022}^{+0.00000022}$ \\
             
             $T_{0}$ [BJD] & $2458870.692582_{-0.000159}^{+0.000163}$ \\
             
             $b$ & $0.358_{-0.033}^{+0.030}$ \\
             
             $u_{0}$ & $0.187_{-0.113}^{+0.128}$ \\
             
             $u_{1}$ & $0.507_{-0.214}^{+0.196}$ \\
            
             $\left(R_{p}/R_{s}\right)_{21}$ & $0.01708_{-0.00024}^{+0.00024}$ \\
             $\left(R_{p}/R_{s}\right)_{44}$ & $0.01684_{-0.00024}^{+0.00023}$ \\
             $\left(R_{p}/R_{s}\right)_{46}$ & $0.01675_{-0.00025}^{+0.00023}$ \\
             
             $\delta_{t, 21}$ [ppm] & $337_{-8.3}^{+8.3}$ \\
             $\delta_{t, 44}$ [ppm] & $328_{-7.9}^{+8.0}$ \\
             $\delta_{t, 46}$ [ppm] & $324_{-8.4}^{+7.7}$ \\
             
             $\delta_{e, 21}$ [ppm] & $15.40_{-4.11}^{+4.11}$ \\
             $\delta_{e, 44}$ [ppm] & $0.25_{-4.33}^{+4.33}$ \\
             $\delta_{e, 46}$ [ppm] & $7.86_{-4.56}^{+4.57}$ \\
             
            \hline
        \end{tabular}
        \label{tab:bestfit}
\end{table}

\subsection{Occultation depth}
\label{subsection:eclipse}

\begin{figure*}%
    \centering
    \resizebox{\hsize}{!}{\includegraphics{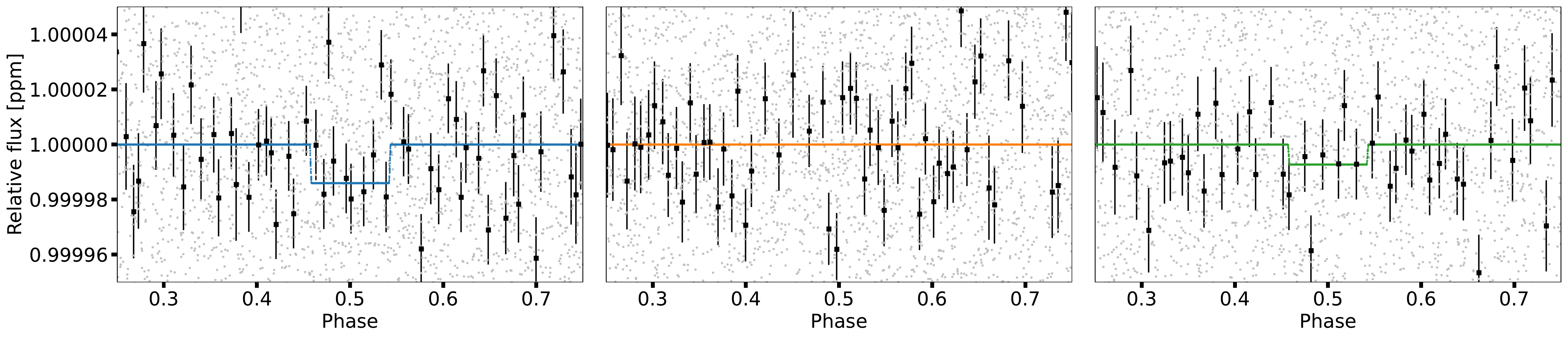}}
    \caption{Zoom on the phase folded relative flux displaying the occultation. The silver dots are detrended flux, black dots are binned data with its corresponding uncertainty. \textit{Left}: Detrended flux and best-fit occultation model in blue, corresponding to sector 21. \textit{Middle}: Detrended flux and best-fit occultation model in orange, corresponding to sector 44. \textit{Right}: Detrended flux and best-fit occultation model in green, corresponding to sector 46.}%
    \label{fig:eclipse panel}%
\end{figure*}

\begin{figure}
    \centering
    \resizebox{0.8\hsize}{!}{\includegraphics{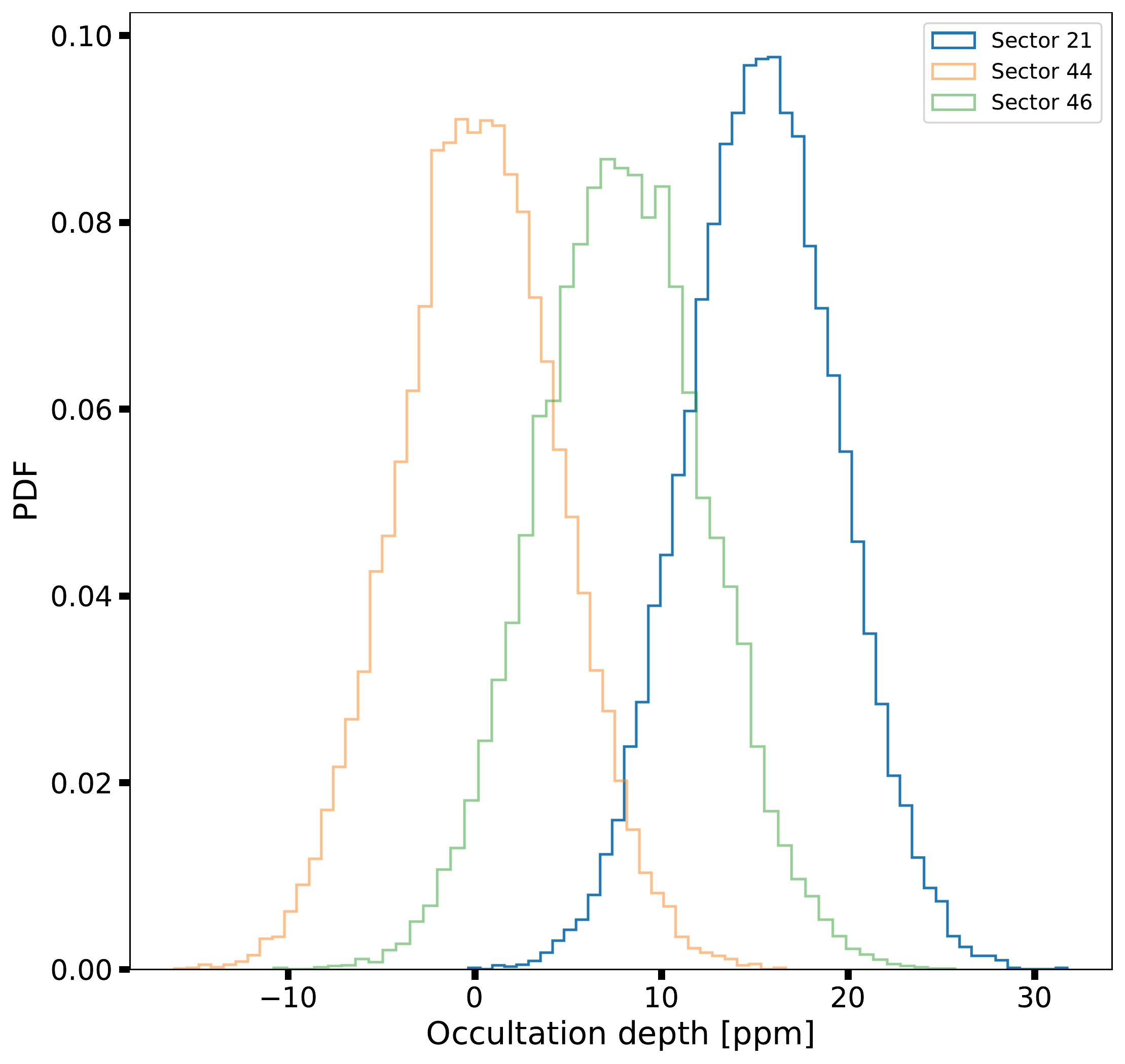}}
    \caption{Posterior density distribution functions for the occultation depth parameter in sector 21 (blue), sector 44 (orange) and sector 46 (green) in the lightcurve model.}
    \label{fig:occultation}
\end{figure}

The occultation depths for each sector are shown in Table~\ref{tab:bestfit}, the resulting occultation lightcurves for each sector are shown in Fig. \ref{fig:eclipse panel} and the posterior distributions in Fig. \ref{fig:occultation}. The correlation between free parameters and its posterior distributions are shown in Fig. \ref{fig:eclipse corner}. Our composite occultation model yields a depth of $8.11_{-2.50}^{+2.62}$ ppm, confirming a positive detection for the complete TESS dataset. The occultation depth of sector 21 is consistent within 1$\sigma$ of the value reported by \citet{Kipping_2020}. Between sector 21 and 44 there is a significant 3.4$\sigma$ decrease in the occultation and increasing marginally 1.6$\sigma$ from sector 44 to 46.

\begin{figure*}%
    \centering
    \resizebox{\hsize}{!}{\includegraphics{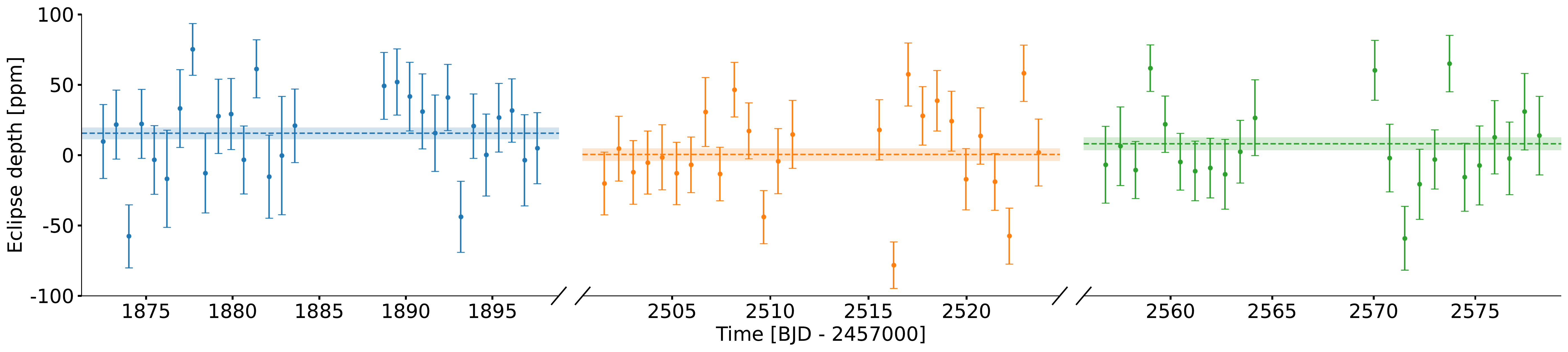}}
    \caption{Individual occultation depth vs. time. Partial occultations are discarded. Blue circles correspond to the best-fit median value in sector 21, orange circles belong to measurements in sector 44 and green circles to sector 46. The errorbars represent 1$\sigma$ uncertainty. Horizontal dashed lines and shaded area are the best-fit median value and 1$\sigma$ uncertainty obtained by fitting over the entire sector (see Table \ref{tab:bestfit}).}%
    \label{fig:alleclipses}%
\end{figure*}

Our depths are considerably smaller compared to the occultation depth measured in the mid-infrared with Spitzer \citep{Demory_2015}. During the 2012 and 2013 campaigns, they measured an occultation depth of $47 \pm 21$ ppm and $176 \pm 26$ ppm, respectively. This difference is expected due to the stronger thermal emission of planet e in the Spitzer bandpass than in TESS.

\begin{figure*}%
    \centering
    \resizebox{\hsize}{!}{\includegraphics{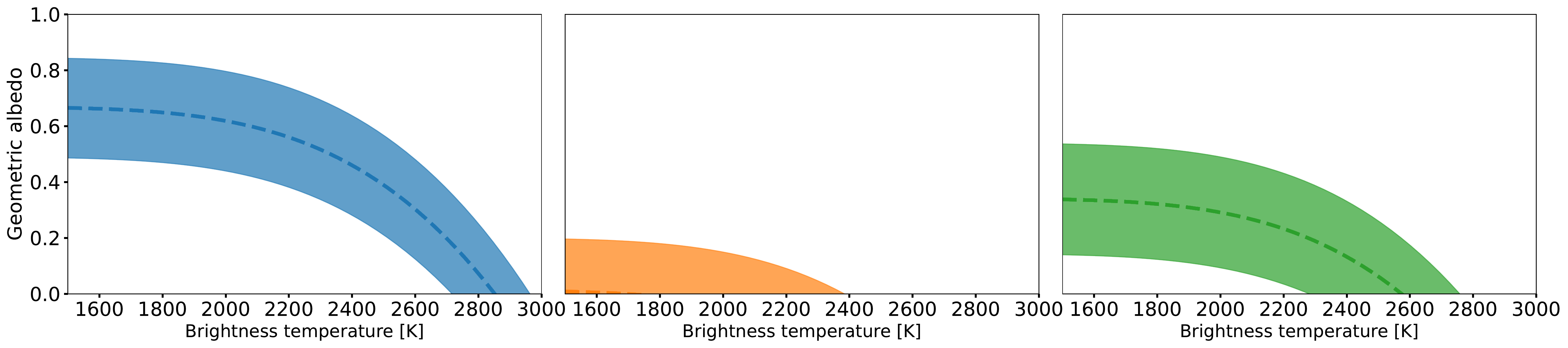}}
    \caption{Relationship between geometric albedo and brightness temperature constrained by the TESS measurements. The dashed line shows the contour for the measured occultation depth for each sector, and the shaded area represents 1$\sigma$ confidence interval based on the results in Table \ref{tab:bestfit}. \textit{Left}: Sector 21. \textit{Middle}: Sector 44. \textit{Right}: Sector 46.}%
    \label{fig:albedo panel}%
\end{figure*}

\subsection{Reflected light and thermal emission}
\label{section:albedo}

To put our results into perspective, we estimate the thermal contribution in the TESS bandpass. We retrieve a theoretical stellar spectrum from PHOENIX stellar model \citep{Husser_2013} with an effective temperature of 5200 K, surface gravity $log(g)=4.5$ \citep{VonBraun_2011} and a planet temperature of 2697 K, which is the maximum hemisphere-averaged temperature measured by \citet{Demory_2016b} with Spitzer observations. Given these values, the thermal contribution in the TESS bandpass is 10.75 ppm. Thus, the occultation depths in sector 21 and 46 are compatible with the thermal contribution within 1$\sigma$, while the depth in sector 44 is approximately 2$\sigma$ below this value. 

For a given occultation depth, we estimate the possible contribution of the reflected light. The geometric albedo $A_{g}$ can be related to the thermal emission and reflected component as \citep{Mallonn_2019}

\begin{IEEEeqnarray*}{rll}
\label{eqn:albedo}
    &A_{g}& = \delta\left(\frac{a}{R_{p}}\right)^2-\frac{B(\lambda, T_{p})}{B(\lambda, T_{s})}\left(\frac{a}{R_{s}}\right)^2 \IEEEyesnumber
,\end{IEEEeqnarray*}

where $d$ is the measured occultation depth, $a$ is the orbital semi-major axis, $R_{p}$ and $R_{s}$ are the planetary and stellar radius, respectively; $B(\lambda, T_{p})$ and $B(\lambda, T_{s})$ are the blackbody emissions of planetary day-side and the star at temperatures $T_{p}$ and $T_{s}$, respectively. Using Eq. \ref{eqn:albedo} we derive the geometric albedo for a range of brightness temperatures of planetary day-side between 1500 K and 3000 K. The possible contributions of reflected light and thermal emission given the occultation depths in the TESS bandpass are shown in Figure \ref{fig:albedo panel}. In each panel, the curve is the contour for the measured occultation depth for a corresponding sector. The brightness temperature represents the thermal emission, while the geometric albedo represents the reflected light \citep{Demory_2011b}. The geometric albedo and brightness temperature estimates are biased because the baseline planet flux is unknown, given it is varying. If we assume that the 4.5 $\mu$m Spitzer measurements can be extrapolated to the TESS wavelengths and adopt the maximum hemisphere-averaged temperature of 2697 K derived by \citet{Demory_2016b}, we infer an upper limit of 0.379 for the geometric albedo.

\section{Discussion}
\label{section:discussion}

From our analysis we draw several conclusions. First, the transit depth across sectors is consistent within the uncertainties. Second, from the combined observations, we measure an occultation depth of $8 \pm 2.5$ ppm. And finally, the occultation varies from sector to sector, from 1.6$\sigma$ to 3.4$\sigma$ significance.

To study how significant our results are, we compare our models by measuring the relative likelihood to describe the observations while penalizing the number of parameters with the Leave-One-Out (LOO) Cross-Validation statistic, as done in \citet{Morris_2021}. In general, the preferred model is ranked first with a $\Delta$LOO of zero. More significant preference for a model relative to another yields a higher $\Delta$LOO \citep{Vehtari_2015,Vehtari_2016}. The weight of a model can be interpreted as the probability to perform best with future data among the considered models \citep{Yao2018}. 

The results are summarised in Table \ref{tab:comparison}. The varying occultation model is preferred, followed by the combined occultation model. The model with no occultation is ranked last. The models including an occultation as parameter have a combined weight of 0.926, which strengthens the evidence of a positive detection. Moreover, the model with an occultation parameter for each sector takes most of the weight, being the one with more chances to perform best on future observations. From our MCMC best-fit and model comparison, we conclude that given the TESS observations, the occultation is detected and slightly favours a variable depth.

\begin{table}[h]
        \centering\setstretch{1.5}
        \caption{Difference in Leave-One-out Cross-Validation and weight between our model fitting an individual occultation per sector (labelled \textit{Occultation per sector}), a composite model of an occultation for all observations together (\textit{Combined occultation}) and a model with no occultation (\textit{No occultation}).}
        \begin{tabular}{c | c c c}
            Model & Rank & $\Delta$ LOO & Weight \\
            \hline 
             Occultation per sector & 1 & 0 & 0.639 \\
             
             Combined occultation & 2 & 2.61 & 0.287 \\
            
             No occultation & 3 & 11.38 & 0.074 \\

            \hline
        \end{tabular}
        \label{tab:comparison}
\end{table}

To get a better sense of the occultation variability, we use our MCMC algorithm to estimate the depth for each individual occultation. We discard observations of partial occultations given the small number of measurements. The result is shown in Fig. \ref{fig:alleclipses}. Negative depths have no physical meaning. The power spectrum on the results do not show a strong periodicity.

Considering the evidence provided in this study alone, the process responsible for a change in occultation depth remains unknown. If the change in occultation is of astrophysical origin, the planet undergoes a process that interchangeably obscures and brightens either its surface or the close vicinity of the planet. It is possible that TESS observed the system at different levels of activity (e.g. volcanism) in each sector. As mentioned in Sect. \ref{section:introduction} the variability could be of stellar origin, an effect of star-planet interaction, catastrophic disintegration, change in opacity due to volcanic activity, due to the presence of an inhomogeneous circumstellar torus of dust or another unidentified process.

Given planet e's extremely short period, it is natural to compare it with other Ultra Short Period (USP) planets. Due to the strong stellar irradiation, Mercury-size planets can evaporate and lead to disintegration \citep{Rappaport_2012}. However, based on our evidence we rule out an asymmetric transit shape (see Fig. \ref{fig:transit}), characteristic of a disintegrating planet due to a tail (and possibly leading trail) of material, such as the case of KIC~12557548. Moreover, the residuals do not exhibit an excess or depression of light relative to the mean out-of-transit flux shortly before ingress or after egress. 

Since our measurements point towards a constant transit depth but a variable occultation depth, it is possible that the planet or its vicinity is covered by a variable amount of material with significant back-scattering and little forward scattering \citep{Sanchis-Ojeda_2015}.     

\begin{figure*}
\centering
\subfigure{\includegraphics[width=0.32\textwidth]{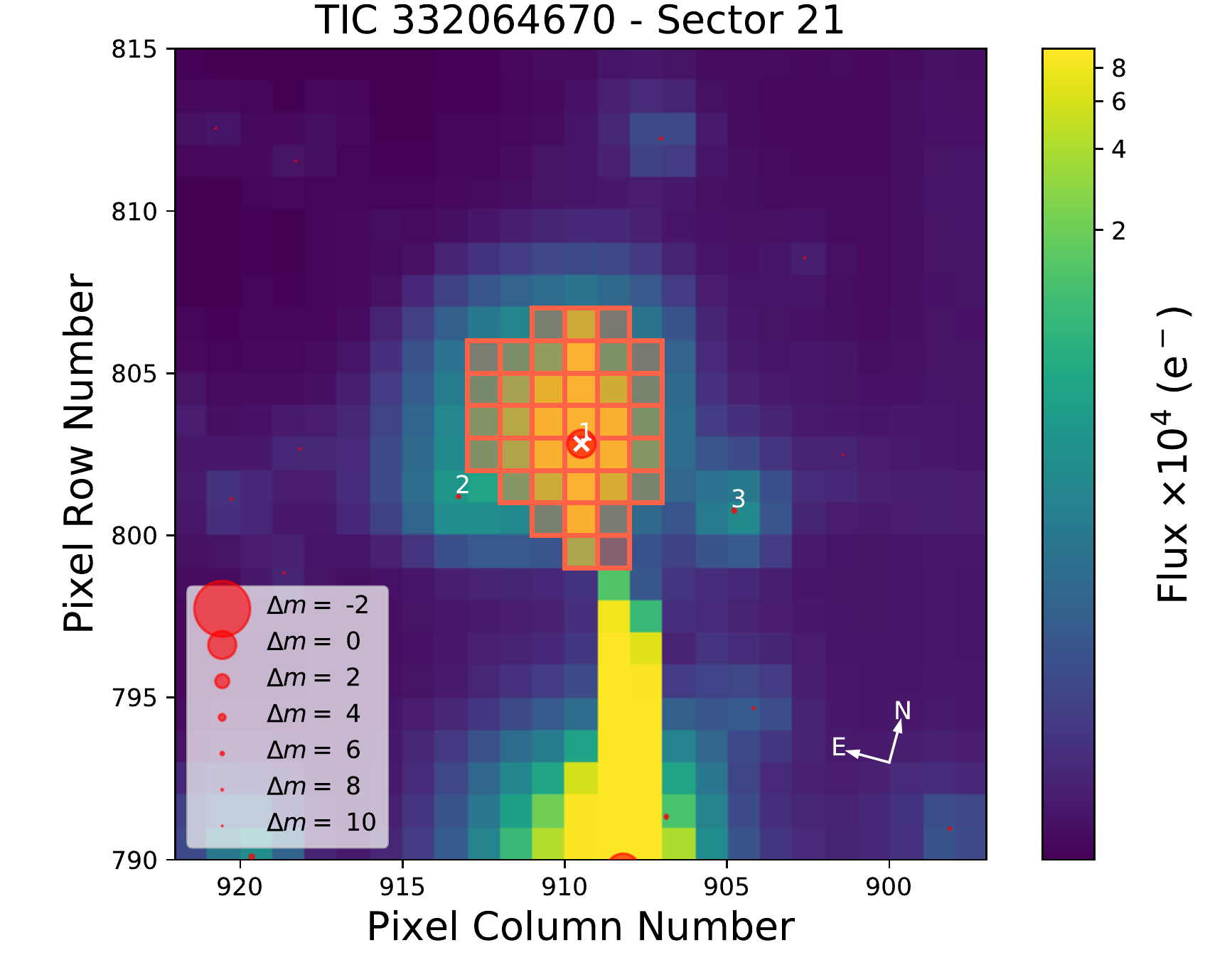}} 
\subfigure{\includegraphics[width=0.32\textwidth]{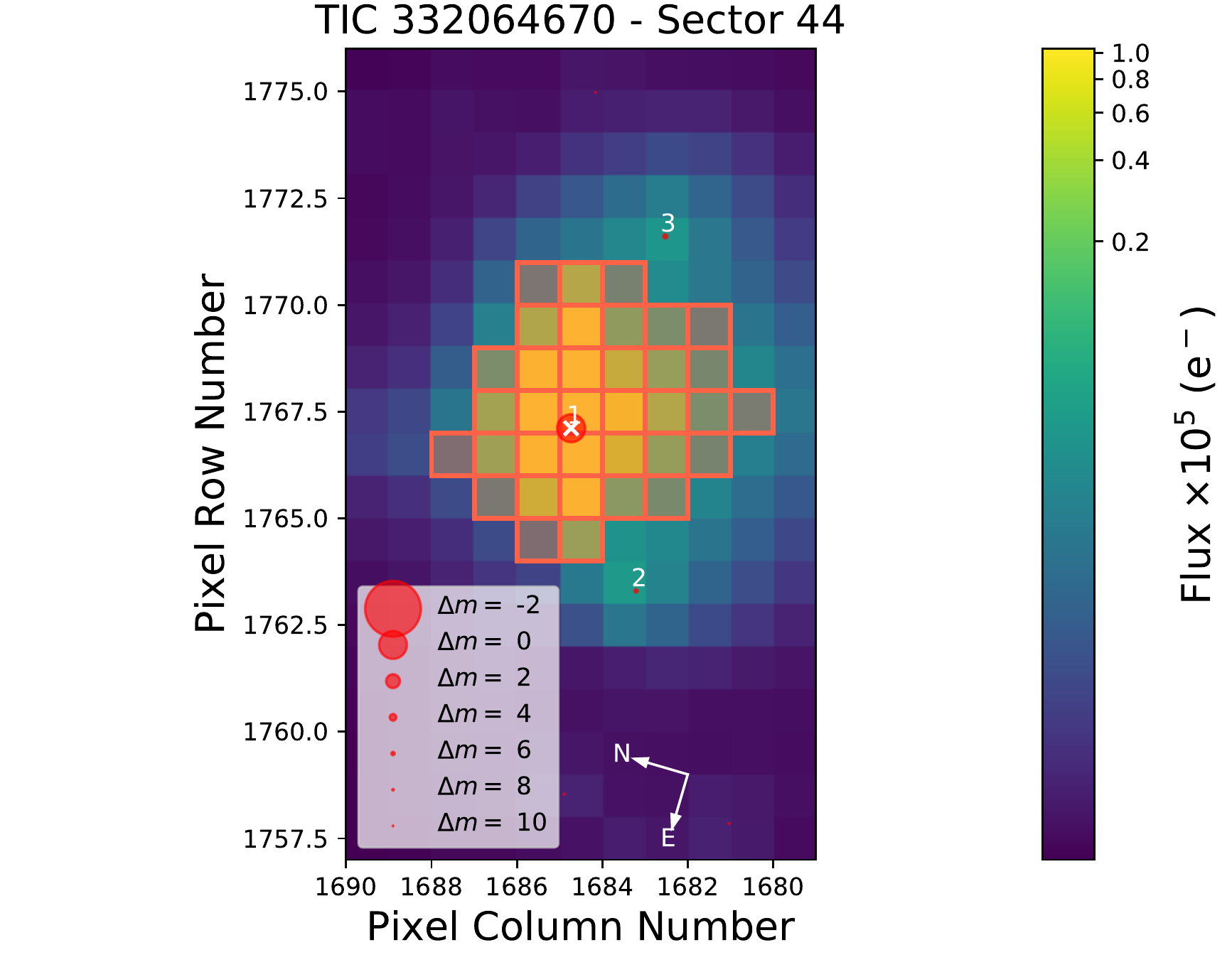}} 
\subfigure{\includegraphics[width=0.32\textwidth]{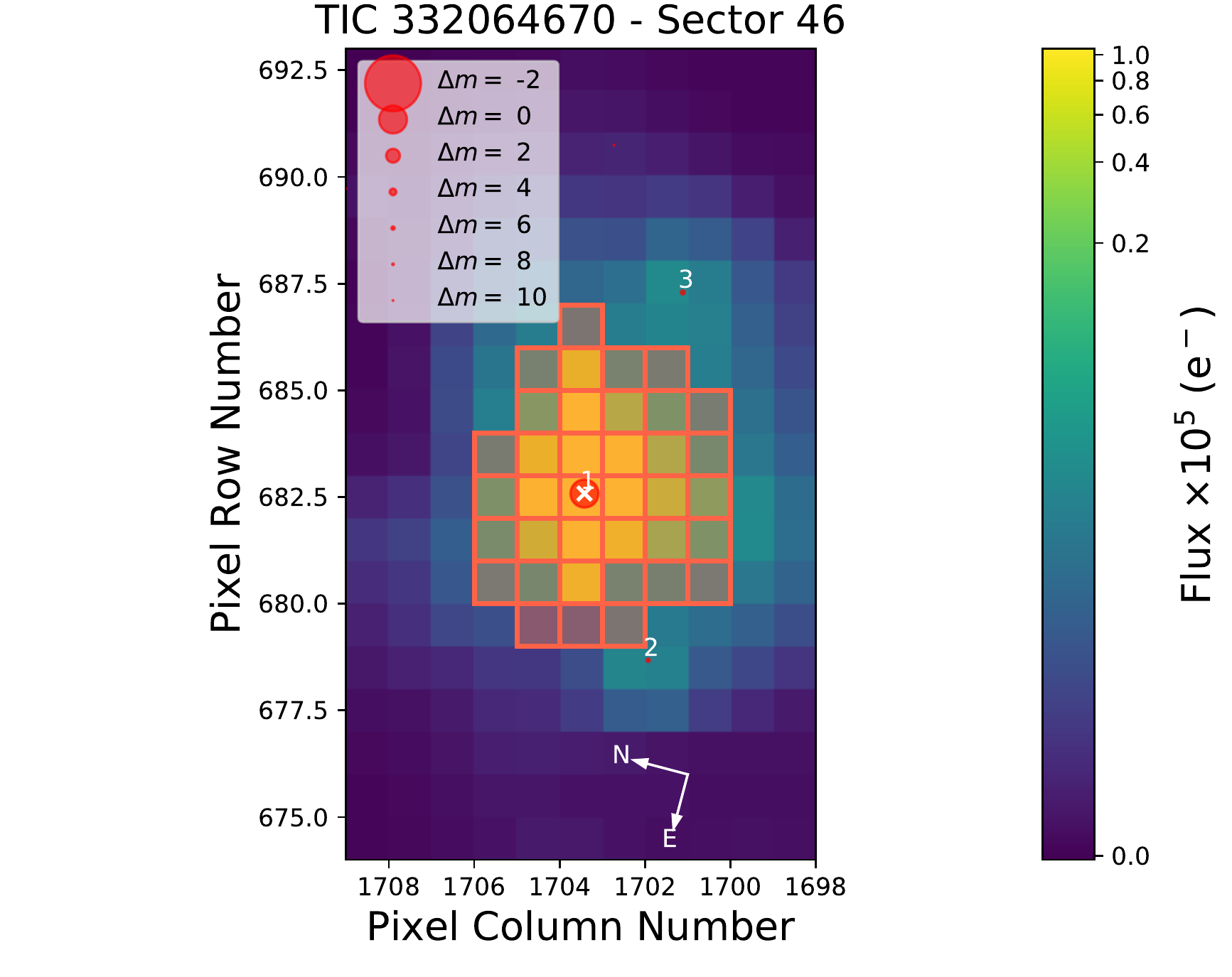}} 
\caption{Target pixel file (TPF) images for 55 Cnc, from TESS sectors 21 (left), 44 (centre) and 46 (right). The apertures used for lightcurve generation are over-plotted in red contour lines; sources identified in Gaia DR2 are also included, with symbols correlated to their brightness compared to the target. Star number 2 is the companion, 55 Cnc B, while the much brighter 53 Cnc is seen at the bottom of the image for sector 21. Star number 3 can be found in the Gaia EDR3 \citep{Gaia_2016, Gaia_2020} under the ID 704966693493530496.  
These images were produced using \texttt{tpfplotter} \citep{Aller_2020}}
\label{fig:tpfs}
\end{figure*}

Variable contamination across different sectors could change the detected occultation depth, and we note that the orientation of the spacecraft was different in sector 21 than the later two sectors. Inspecting the TPFs (see Fig. \ref{fig:tpfs}) reveals that the contamination by 53 Cnc is minimal in the latter two sectors, and may affect sector 21. However, the occultation depth in sector 21 is the largest, and contamination would bias the occultation towards shallower depths, so we infer that removing the contamination would only strengthen the detection in sector 21.

As validation of our derived uncertainties in the occultation depth, we perform injection tests, which consist in injecting mock transits and occultations in the lightcurve residuals. We construct the synthetic lightcurve with \texttt{batman} \citep{Kreidberg_2015}. The time at mid-transit was chosen randomly between [$t_{0}+T_{14}/2, t_{0}+(P-T_{14})/2$] (somewhere between end of true transit and start of occultation), where $t_{0}$ is 55 Cnc e's mid-transit time, $T_{14}$ its transit duration and $P$ its period. Based on the thermal contribution computed in Sect. \ref{section:albedo}, we choose to inject an occultation of 10 ppm in our residuals. Then we repeat the same exercise but in a randomly drawn mid-transit time between [$t_{0}+(P+T_{14})/2, t_{0}+P-T_{14}/2$] (between end of true occultation and start of transit).

The same MCMC algorithm as described in Sect. \ref{section:method} is used on the data for one sector at a time. For each sector individually, we find a mid-transit time and occultation depth agreeing with the corresponding true values within 1$\sigma$. The uncertainties seem to be comparable to our results in Sect. \ref{section:results}, but in general tend to over- or underestimate, pointing at correlated noise still present in the lightcurve.
    
\section{Conclusions}
\label{section:conclusions}

At the present stage, we know that the planet exhibits a phase modulation too large to be attributed to reflected light and thermal emission \citep{Winn_2011, Sulis_2019} and undergoes a significant change in day-side brightness temperature over time \citep{Demory_2015, Tamburo_2018}. So far, the variability in the occultation depth has only been observed in the IR. 
Here we confirm the detection of the occultation on the combined TESS observations and present weak evidence of a variable occultation in the optical. The process causing these phenomena is still unknown. Based on our results, possible contribution of reflected light in the measured signal put an upper limit of 0.4 on the geometric albedo.

The exquisite precision demonstrated in CHEOPS observing this system \citep{Morris_2021} and the much anticipated JWST will most likely provide exciting findings about this enigmatic system. In particular, two proposals to observe 55 Cnc e in Cycle 1 were accepted. One program aims to identify if the origin of the variable occultation depth is due to a 3:2 spin-orbit resonance \citep{Brandeker_2021}, resulting in a different side of the planet visible. The second project focuses on atmospheric characterization by measuring the thermal emission spectrum from 3.8-12 micron \citep{Hu_2021}. Furthermore, the planet K2-141b, a so-called lava world and similar in characteristics to 55 Cnc e, will also be observed by JWST \citep{Dang_2021}. In the coming months, we might not only learn more about 55 Cnc e's nature, but about the USP population in general.

\begin{acknowledgements}
We are grateful to the anonymous referee for the thoughtful comments that improved this paper. EMV thanks A. Oza and H. Osborn for helpful discussions. 
This work has received support from the Centre for Space and Habitability (CSH) and the National Centre for Competence in Research PlanetS, supported by the Swiss National Science Foundation (SNSF).
RW, NS and B.-O. D. acknowledges support from the Swiss National Science Foundation (PP00P2-190080).
This paper includes data collected by the TESS mission. Funding for the TESS mission is provided by the NASA's Science Mission Directorate.
This research made use of \texttt{exoplanet} \citep{Foreman-Mackey_2021} and its
dependencies \citep{exoplanet:agol20, exoplanet:arviz, exoplanet:astropy13,
exoplanet:astropy18, exoplanet:kipping13, exoplanet:luger18, Salvatier_2016,
exoplanet:theano}. This research made use of \texttt{Lightkurve}, a Python package for Kepler and TESS data analysis \citep{lightkurve_2018}. We acknowledge the use of further software: \texttt{NumPy} \citep{Harris_2020}, \texttt{matplotlib} \citep{Hunter_2007}, \texttt{corner} \citep{corner_2016}, \texttt{astroquery} \citep{Ginsburg_2019} and \texttt{scipy} \citep{scipy_2020}.
This work has made use of data from the European Space Agency (ESA) mission
{\it Gaia} (\url{https://www.cosmos.esa.int/gaia}), processed by the {\it Gaia}
Data Processing and Analysis Consortium (DPAC,
\url{https://www.cosmos.esa.int/web/gaia/dpac/consortium}). Funding for the DPAC
has been provided by national institutions, in particular the institutions
participating in the {\it Gaia} Multilateral Agreement.
\end{acknowledgements}

\bibliographystyle{aa}
\setcitestyle{authoryear,open={(},close={)}}
\bibliography{reference}

\begin{appendix}
\section{Posterior distributions}
\label{appendix:corner}

\begin{figure*}%
    \centering
    \resizebox{\hsize}{!}{\includegraphics{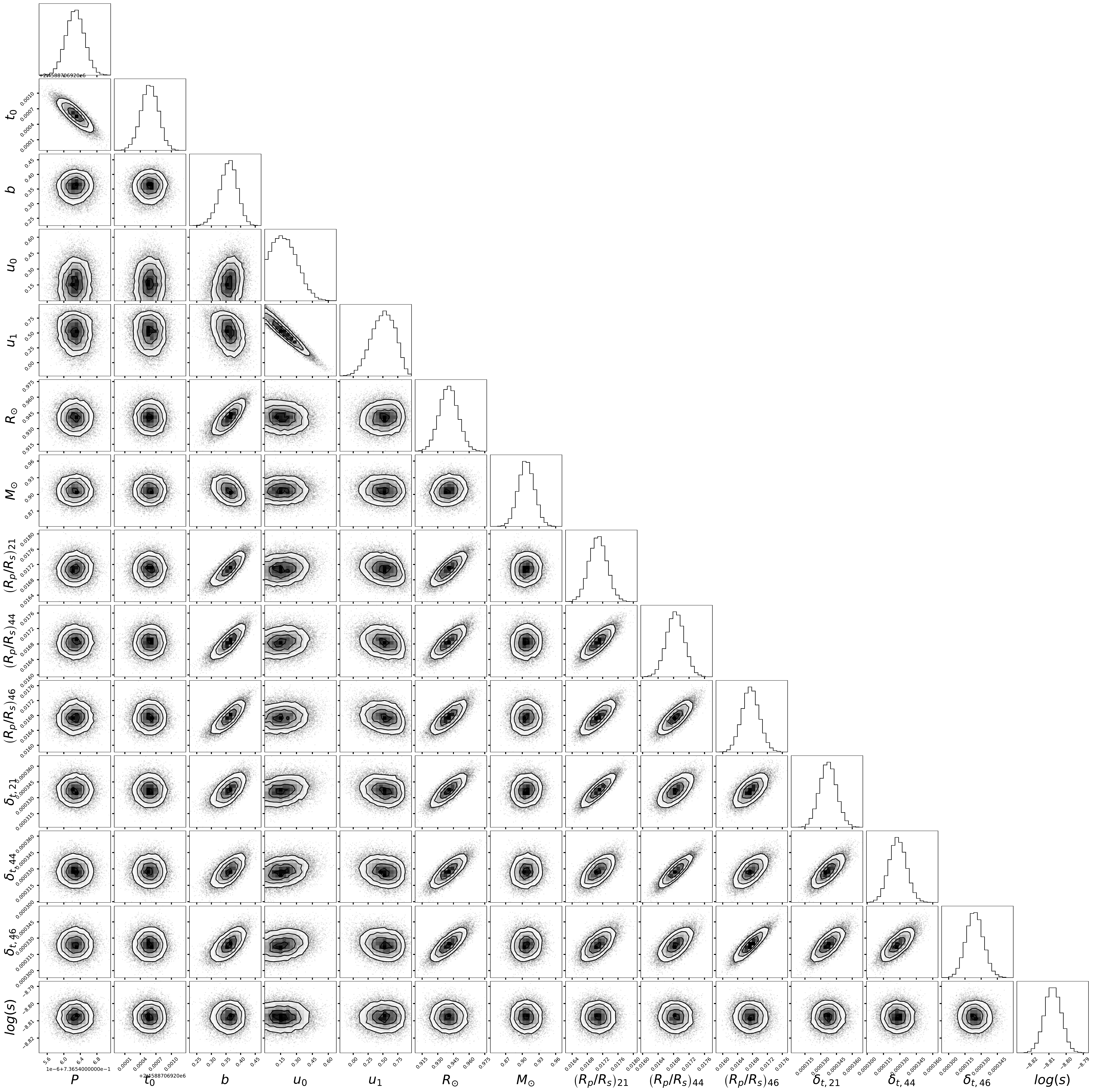}}
    \caption{Posterior distributions and joint correlations between all free parameters in the transit model fit. The parameters are: Orbital period $P$, mid-transit time $t_{0}$, impact parameter $b$, quadratic limb darkening coefficients $u_{0}$ and $u_{1}$, stellar radius $R_{\odot}$ and mass $M_{\odot}$, planet-to-star radius ratio $R_{p}/R_{s}$ and transit depth $\delta_{t}$ with a numeric subscript corresponding to the sector. $log(s)$ is the natural logarithm of the flux uncertainty for each measurement.}%
    \label{fig:transit corner}%
\end{figure*}

\begin{figure*}%
    \centering
    \resizebox{0.7\hsize}{!}{\includegraphics{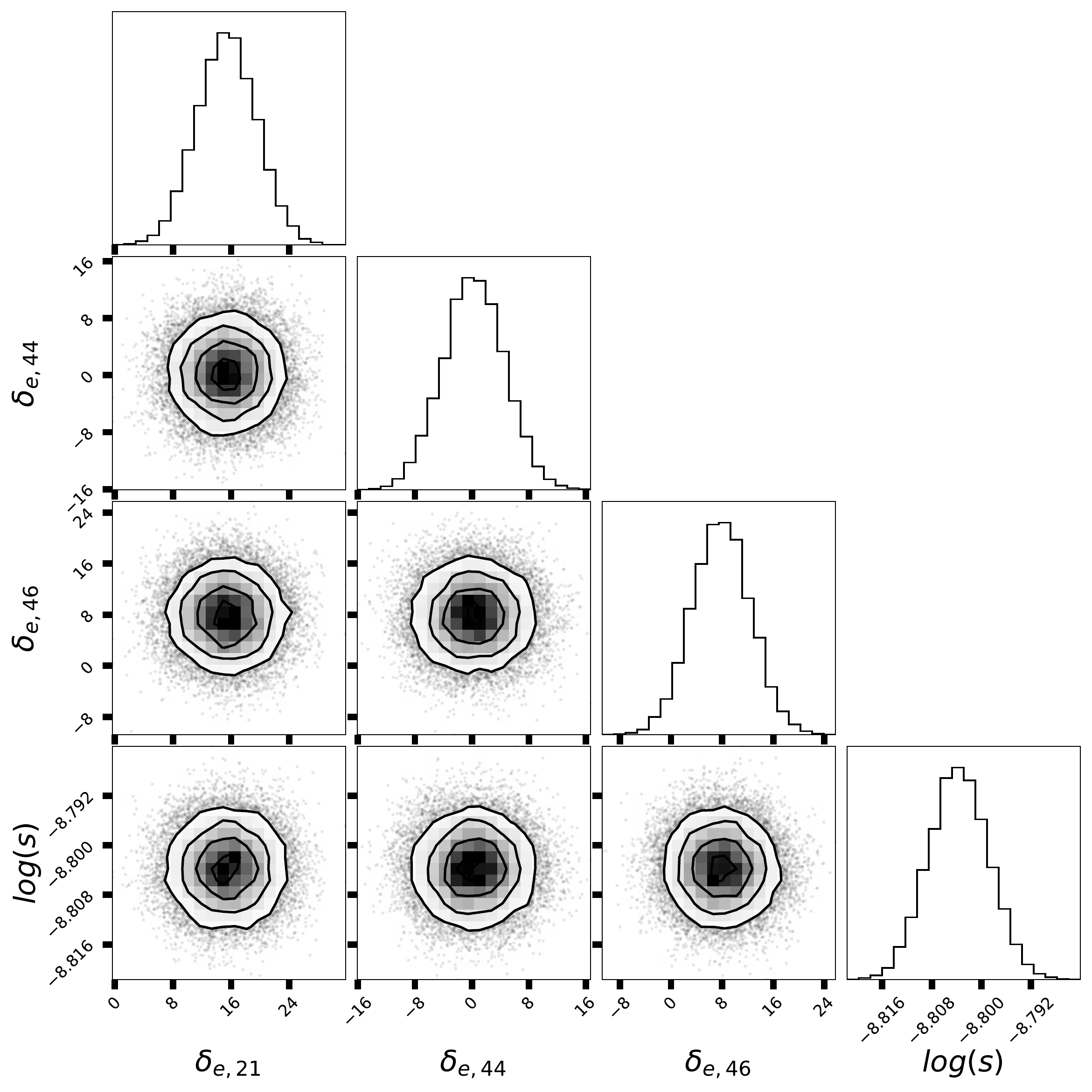}}
    \caption{Posterior distributions and joint correlations between all free parameters in the occultation model fit. The parameters shown are: Occultation depth $\delta_{t}$ with a numeric subscript corresponding to the sector. $log(s)$ is the natural logarithm of the flux uncertainty for each measurement.}%
    \label{fig:eclipse corner}%
\end{figure*}

For completeness, we present the corner plot of the parameters sampled from the transit model MCMC fit in Fig. \ref{fig:transit corner}, while Fig. \ref{fig:eclipse corner} presents some parameters sampled from the occultation model MCMC fit.

\section{RMS vs. bin size}
\label{appendix:rms}

\begin{figure*}%
    \centering
    \resizebox{\hsize}{!}{\includegraphics{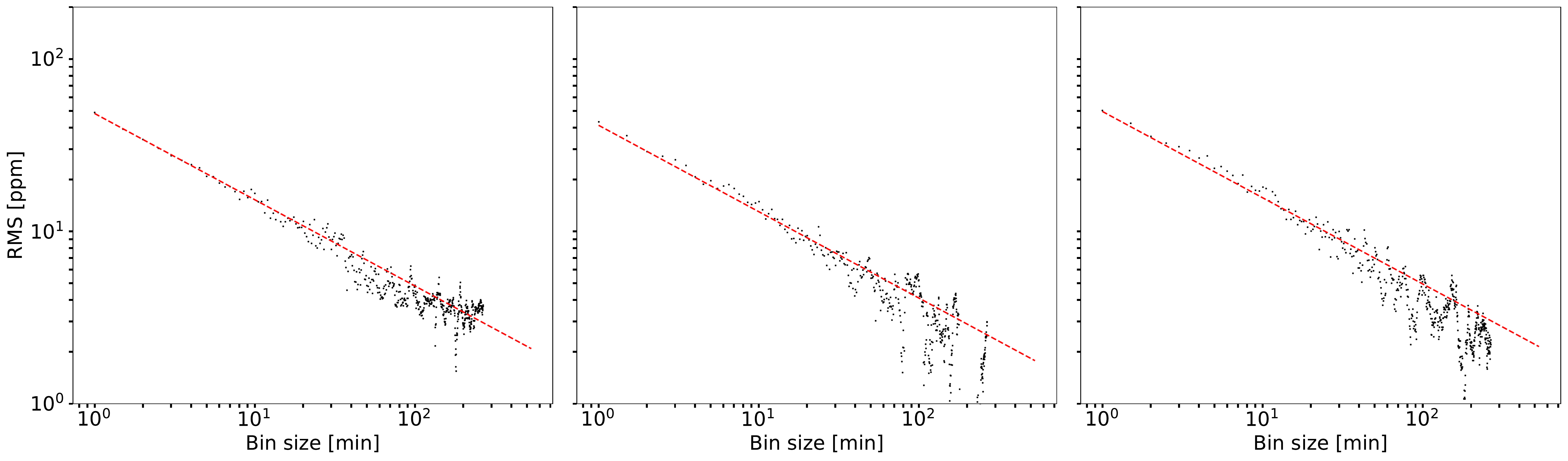}}
    \caption{Photometric residual RMS for different bin size in minutes for sector 21 (left), sector 44 (middle) and sector 46 (right). The red dashed line shows the expected decrease in Poisson noise precision normalised to a 2-min bin.}%
    \label{fig:rms}%
\end{figure*}

If the remaining noise in the observations is white, the residual RMS should decrease as 1/$\sqrt(n)$, where $n$ is the size of the bin. The resulting plots of our occultation model residuals are shown in Fig. \ref{fig:rms}.

\end{appendix}

\end{document}